\documentstyle[prb,aps,psfig]{revtex}
\begin{document}

\draft
\twocolumn[
\widetext

\title{Transverse Pseudospin Susceptibility and Tunneling Parameters of
Double Layer Electron Gas Systems}
\author{L. \'Swierkowski}
\address{School of Physics, The University of New South Wales,
Sydney 2052, Australia}
\author{A.H. MacDonald}
\address{Department of Physics, Indiana University,
Bloomington IN 47405, USA}

\date{\today}
\maketitle

\widetext
\begin{abstract}
\leftskip 2cm
\rightskip 2cm
The subbands of weakly coupled double-layer two dimensional electron
gas systems consist of narrowly spaced pairs whose corresponding
wavefunctions are symmetric and antisymmetric combinations of
isolated layer subband wavefunctions.  The energetic spacing within a
pair is $2t$ where $t$ is the interlayer tunneling amplitude.
$t$ is an important parameter in modeling these systems and, if
interactions could be neglected, it would be  proportional to beating
frequencies seen in weak-field magnetic oscillations experiments and
therefore readily measurable. We point out that interactions alter the
beating frequency.  We discuss similarities and differences between this
effect and exchange-correlation enhanced spin-splitting.
\end{abstract}
\pacs{\leftskip 2cm PACS numbers:  73.20.Dx, 73.40.Hm, 75.20.En}
]
\narrowtext

The development of of high mobility double-quantum-well structures has
opened up a fruitful area of research in low-dimensional physics.  The
many-body physics of these systems is especially interesting in the strong
magnetic field regime where quenching  of the kinetic energy due to Landau
quantization leads to a variety  of strong correlation
effects\cite{dassarmabook}, including novel quantum Hall
effects\cite{dlqhexp,dlqhtheory} and spontaneous interlayer phase
coherence.\cite{coherenceexp,coherenceth} In models of  these systems a
critical parameter is the amplitude $t$ for quantum tunneling of electrons
between the layers.   In a tight-binding approximation, the 
subbands in weakly coupled electrically balanced double-layer
systems occur in pairs, separated in energy by $2 t$, 
whose wavefunctions are the  symmetric and antisymmetric combinations of
isolated layer subband wavefunctions.  For individual samples the
parameter $t$  is routinely\cite{dlqhexp,coherenceexp,wfexp,shayegan}
determined by analyzing Shubnikov-de Haas oscillations in the resistance
at weak magnetic fields.  The magnetic oscillation frequencies are
proportional to  the areas enclosed by the circular Fermi circles for the two
subbands.  In the past it has  been assumed that for double-layer systems the
difference in cross-sectional areas of the symmetric and antisymmetric subbands 
can be related to $t$ by using the following equation which would be valid for 
non-interacting electrons: 
\begin{equation}
\frac{n_{S}-n_{A}}{\nu_{0}} = 
\frac{\hbar^{2}( k_{FS}^{2}-k_{FA}^{2})}{2m^{*}} = 2t .
\label{eq:1}
\end{equation}
Here $k_{F}$ and $n$ are the Fermi wavevectors and densities for
symmetric ($S$) and antisymmetric ($A$) subbands, $m^{*}$ is the
semiconductor band structure effective mass, and $\nu_{0} =
m^{*}/(\pi\hbar^{2})$ is the non-interacting electron density of states
for each two-dimensional subband.   In this paper we point out that when
electron-electron  interactions are included Eq.~\ref{eq:1} is no longer
valid.  The interlayer tunneling parameter appearing this equation must
be replaced by an effective value $t^{*}$. The change in the effective
value of $t$ is analogous to the familiar exchange-correlation 
enhancement of the spin-splitting of electronic bands in the presence of a
magnetic field.  However, as we will discuss,  there are important
differences.  As a  consequences of these, the effective splitting can be
either enhanced or suppressed depending on the layer separation.

We will couch our analysis in terms of a pseudospin
representation\cite{oldpaper}  for the layer degree of freedom which has
proved useful in work on double-layer systems in strong magnetic fields.
We restrict our attention here to the case where the two coupled layers are
identical.  Then the second quantized Hamiltonian of the system $H
= H_{1} + H_{2}$, where the one body term
\begin{equation}
H_{1} = \sum_{\vec{k},s,\sigma} \left[\frac{\hbar^{2}k^{2}}{2m^{*}}
c^{\dagger}_{\vec{k},s,\sigma}
c^{\phantom{\dagger}}_{\vec{k},s,\sigma}- t\;
c^{\dagger}_{\vec{k},s,\sigma}
c^{\phantom{\dagger}}_{\vec{k},s,\bar{\sigma}} \right]  
\label{eq:2}
\end{equation}
and the interaction term
\begin{eqnarray}
H_{2} &=& \frac{1}{2} \sum_{\vec{k},s,\sigma} \sum_{\vec{k}',s',\sigma'}
\sum_{\vec{q}} \left[V_{0}(\vec{q}) + \sigma \sigma'
V_{z}(\vec{q})\right]\nonumber\\
 && c^{\dagger}_{\vec{k}+\vec{q},s,\sigma}
c^{\dagger}_{\vec{k}'-\vec{q},s',\sigma'}
c^{\phantom{\dagger}}_{\vec{k}',s',\sigma'}
c^{\phantom{\dagger}}_{\vec{k},s,\sigma}.
\label{eq:3}
\end{eqnarray}
Here $s$ in the spin label, $\sigma$ is the pseudospin label,
$V_{0}(\vec{q}) = \left(V_{A}(\vec{q}) + V_{E}(\vec{q})\right)/2$, and
$V_{z}(\vec{q}) = \left(V_{A}(\vec{q}) - V_{E}(\vec{q})\right)/2$  where
$V_{A}(\vec{q})$ and
$V_{E}(\vec{q})$ are respectively the intra and inter layer effective
electron-electron interactions.  ($\sigma = \pm 1$ for electrons in left
and right layers and
$\bar{\sigma} = -\sigma$.)  In the narrow layer limit\cite{widthcaveat},
which we use for illustrative calculations below,
$V_{A}(\vec{q}) = 2\pi e^{2}/q $ and $V_{E}(\vec{q}) = \exp{(-qd)}
V_{A}(\vec{q}) $ where $d$ is the layer separation. 

The invariance of the Hamiltonian under reversal of pseudospin labels
guarantees that the
Greens function is diagonal when expressed in the representation of
symmetric and antisymmetric states.  It therefore follows from the
Luttinger theorem\cite{lutth} 
that the relationship between the Fermi circle areas measured in
magneto-oscillation experiments and the densities of symmetric and
antisymmetric electrons is not altered by interactions. However the
relationship between these densities and the hopping parameter $t$, which
we  calculate below, is altered.  We discuss this relationship in terms of
the linear  response of the $x$ component of the pseudospin polarization
to the hopping term in the  Hamiltonian:
\begin{equation}
n_{S} - n_{A}  = 2 \lim_{\vec{q} \to 0} \langle I^{x}(\vec{q})\rangle
\equiv 2 t \left[\lim_{\vec{q} \to 0}
\chi^{xx}(\vec{q},\omega=0)\right]
\label{eq:4}
\end{equation}
where
\begin{equation}
I^{\alpha}(\vec{q}) = \frac{1}{2} \sum_{\vec{k},s,\sigma,\sigma'}
c^{\dagger}_{\vec{k}-\vec{q}/2,s,\sigma}
\tau^{\alpha}_{\sigma,\sigma'}
c^{\phantom{\dagger}}_{\vec{k}+\vec{q}/2,s,\sigma'}.
\label{eq:5}
\end{equation}
In this equation $\tau^{\alpha}$ is a Pauli spin matrix.  Since the
hopping term can be recognized  as Zeeman coupling to the $x$ component of the 
pseudospin, the required calculation would be identical to that for the spin
susceptibility of a two-dimensional electron system\cite{spinsuscep,mccalc} if
the  pseudospin-dependent term in $H_{2}$, which vanishes for $d \to 0$,
were not present.   We will refer to $\chi^{xx}$ as the {\it
transverse} pseudospin susceptibility to distinguish it  from the response
of the $z$ pseudospin component to a bias potential, which is sharply
reduced by screening effects present already at the level of the Hartree
approximation\cite{longitudinal} for this {\it longitudinal} pseudospin
susceptibility. We have evaluated the transverse pseudospin susceptibility
using the generalized random-phase-approximation (GRPA) and find that it
is qualitatively altered by  the pseudospin dependence of the interaction
term in the Hamiltonian. This calculation is outlined in the following
paragraph.  The GRPA has important  deficiencies which we discuss below,
but it is adequate for the qualitative  issues we address.

In the GRPA the transverse pseudospin susceptibility is approximated by
the following  expression:
\begin{equation}
\chi^{xx}(\vec{q},\omega) =  \frac{2}{A} \sum_{\vec{k}} L_{\vec{k}}(q,\omega)
\left[ \frac{f_{\vec{k}+\vec{q}/2} - f_{\vec{k}-\vec{q}/2}}{\hbar\omega -
\epsilon_{\vec{k}+\vec{q}/2} +
\epsilon_{\vec{k}-\vec{q}/2}}\right]
\label{eq:6}
\end{equation}
where $A$ is the two-dimensional (2D) system area and 
$L_{\vec{k}}(\vec{q},\omega)$ is the solution to the following integral equation:
\begin{eqnarray}
\lefteqn{L_{\vec{k}}(\vec{q},\omega) =  1}\nonumber\\
 &&+ \frac{1}{A}
\sum_{\vec{k}'} V_{E}(\vec{k} - \vec{k}') \left[
\frac{f_{\vec{k}'+\vec{q}/2} - f_{\vec{k}'-\vec{q}/2}}{\hbar\omega -
\epsilon_{\vec{k}'+\vec{q}/2} +
\epsilon_{\vec{k}'-\vec{q}/2}}\right] L_{\vec{k}'}(\vec{q},\omega)
\label{eq:7}
\end{eqnarray}
In Eq.~\ref{eq:6} and Eq.~\ref{eq:7}
$\epsilon_{\vec{k}}$ is the Hartree-Fock approximation
quasiparticle energy dispersion relation at $t=0$,
\begin{equation}
\epsilon_{\vec{k}} = \frac{\hbar^{2}k^{2}}{2m^{*}} - \frac{1}{A}
\sum_{\vec{k}'} f_{\vec{k}'} V_{A}(\vec{k} - \vec{k}'),
\label{eq:8}
\end{equation}
and $f_{\vec{k}}$ is the corresponding Fermi occupation factor.  The
starting point for the derivation of Eq.~\ref{eq:6} is the
fluctuation-dissipation  theorem expression for the transverse pseudospin
susceptibility in terms of thermodynamic fluctuations of the $I^{x}$
pseudospin operator:
\begin{equation}
\chi^{xx}(\vec{q},\omega=0) = \frac{-2i}{\hbar A} \int_{0}^{\infty} dt
\left\langle[I^{x}(- \vec{q},t), I^{x}(\vec{q},t=0)]\right\rangle
\label{eq:9}
\end{equation}
The GRPA is most economically derived\cite{no} by making a Hartree-Fock
factorization of the equation of  motion for $\chi^{xx} (\vec{q},t)$.
Alternately, the same expression can be obtained in diagrammatic
perturbation theory by evaluating ladder-sum vertex correction to the
bubble skeleton diagram. In the limit of interest ($\omega=0$ and $\vec{q}
\to 0$) the ratio of Fermi factors and  quasiparticle energy differences in
these equations simplifies to $(m^{HF}/\hbar k_{F})\delta(k - k_{F})$
where $m^{HF}$ is the Hartree-Fock approximation quasiparticle effective
mass allowing.  With this simplification Eq.~\ref{eq:7} can be solved
analytically and the sum in Eq.~\ref{eq:6} performed.  The result is
\begin{equation}
\chi^{xx}(\vec{q} \to 0,\omega=0) = \nu_{HF} L
\label{eq:10}
\end{equation}
where
\begin{eqnarray}
\lefteqn{\frac{\nu_{HF}}{\nu_{0}} = \frac{m^{HF}}{m^{*}}}\nonumber\\
 && = \left[1 + \frac{\nu_{0}}{2} \int_{0}^{2 \pi} \frac{d\phi}{2\pi} 
\cos{(\phi)}V_{A} \left(k_{F}\sqrt{2(1 - \cos{(\phi)})}\right)\right]^{-1},
\label{eq:11}
\end{eqnarray}
and the on shell vertex factor
\begin{equation}
L = \left[1 -  \frac{\nu_{HF}}{2} \int_{0}^{2 \pi} \frac{d\phi}{2\pi} V_{E}
\left(k_{F}\sqrt{2(1 - \cos{(\phi)})}\right)\right]^{-1}
\label{eq:12}
\end{equation}
Eqs. ~\ref{eq:10}, ~\ref{eq:11}, and ~\ref{eq:12} may be combined to
express the transverse pseudospin susceptibility in a form which is
familiar from the electron gas theory of the spin susceptibility and which
form the basis of the following discussion.
\begin{equation}
\chi^{xx}(\vec{q} \to 0,\omega=0) = \nu_{0}S = \nu_{0}/(1 -I)
\label{eq:13}
\end{equation}
where $S$ is the Stoner enhancement factor and the Stoner interaction parameter 
\begin{equation}
I = \frac{\nu_{0}}{2\pi k_{F}} \int_{0}^{2k_{F}} d q \frac{V_{E}(q) + (2
(q/2k_{F})^{2} - 1) V_{A}(q)}{(1 - (q/2k_{F})^{2})^{1/2}}. 
\label{eq:14}
\end{equation}

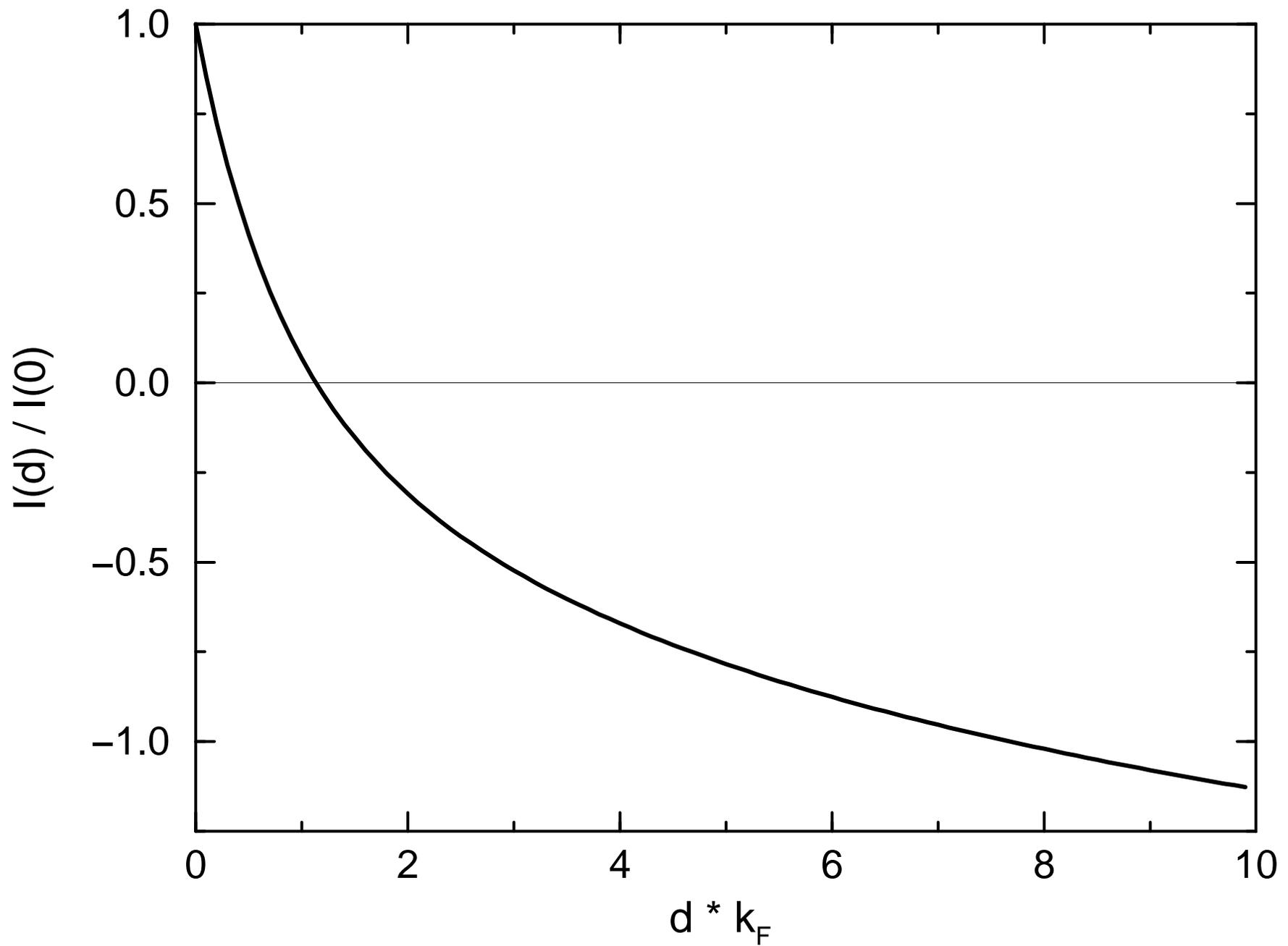
\begin{figure}
\caption{GRPA transverse pseudospin susceptibility Stoner interaction parameter
$I$ as a function of $k_{F} d$.  In this approximation $I(k_F d)/I(d=0)$ is
independent of density.  $I(0) = (2)^{1/2} r_s / \pi$ where $r_s$ is
the usual electron gas density parameter defined here in terms of the density 
per layer, $n = ( \pi r_s^2 a_0^2 )^{-1}$.  Here $a_0 = \hbar^2 / m^{*} e^2$ is 
the semiconductor Bohr radius.  For GaAs $a_0 \sim 10$ nm.  The Stoner 
enhancement parameter $S = (1-I)^{-1}$.}
\label{fig:one}
\end{figure}

The GRPA result for $I$ as a function of layer separation is plotted in
Fig. ~\ref{fig:one}.    The most remarkable feature of this result is that
$I$ changes sign and the  susceptibility is suppressed rather than
enhanced for $k_{F} d > 1.13 $. In an electron gas, the usual spin
susceptibility is invariably enhanced by interactions because the band
energy cost of spin alignment is  partially offset by increased attractive
exchange interactions between parallel spins.  The exchange energy gain
due to the alignment of transverse pseudospin components occurs when coherence
is established between electrons in opposite layers\cite{ldaremark} and is
weaker because it comes only from interlayer interactions.   In the limit
$d \to \infty$, this vertex  contribution to $I$ vanishes and we are left
only with the interaction effect on the quasiparticle dispersion in each
layer.  For the Hartree-Fock approximation
$m^{HF}$ vanishes in Coulombic systems, causing the GRPA value of
$I$ to approach $-\infty$ and $S$ to vanish as $d \to \infty$.   This
property is an artifact of the GRPA.  However the fact that for $d \to
\infty$\, $S$ depends only on the one-particle Greens function is a general 
result.  Using the property that the two layers are uncorrelated we have 
derived a formal general expression for this limit:
\begin{eqnarray}
\lefteqn{\chi^{xx}(\vec{q},\omega) = 2 \int_{- \infty}^{\infty} d\epsilon
\int_{-\infty}^{\infty} d{\epsilon}'
\int \frac{d\vec{k}}{(2 \pi)^{2}}}\nonumber\\
&&  \frac{f({\epsilon}') - f(\epsilon)}{\hbar \omega - {\epsilon}' + \epsilon + 
i \eta }  \; \rho_{\vec{k}+\vec{q}/2}({\epsilon}') 
\rho_{\vec{k}-\vec{q}/2}(\epsilon)
\label{eq:15}
\end{eqnarray}
where $\rho_{\vec{k}}(\epsilon)$ is the spectral
weight function\cite{no} of the isolated layer one-body Greens function
normalized so that it integrates to 
unity.  The quasiparticle pole contribution to
$\lim_{\vec{q} \to 0} \chi^{xx}(\vec{q})$ for $d \to \infty$, which 
we expect to be dominant, gives a Stoner enhancement factor
$ S = z^{2} m^{QP}/m^{*} $  where $z$ is the quasiparticle
normalization factor, and $m^{QP}$ is the quasiparticle  effective mass at
the Fermi energy.  For moderate density 2D electron systems this factor
is considerably smaller than unity\cite{mccalc,ptmstar}, in qualitative 
agreement with the large $d$ GRPA  result. 

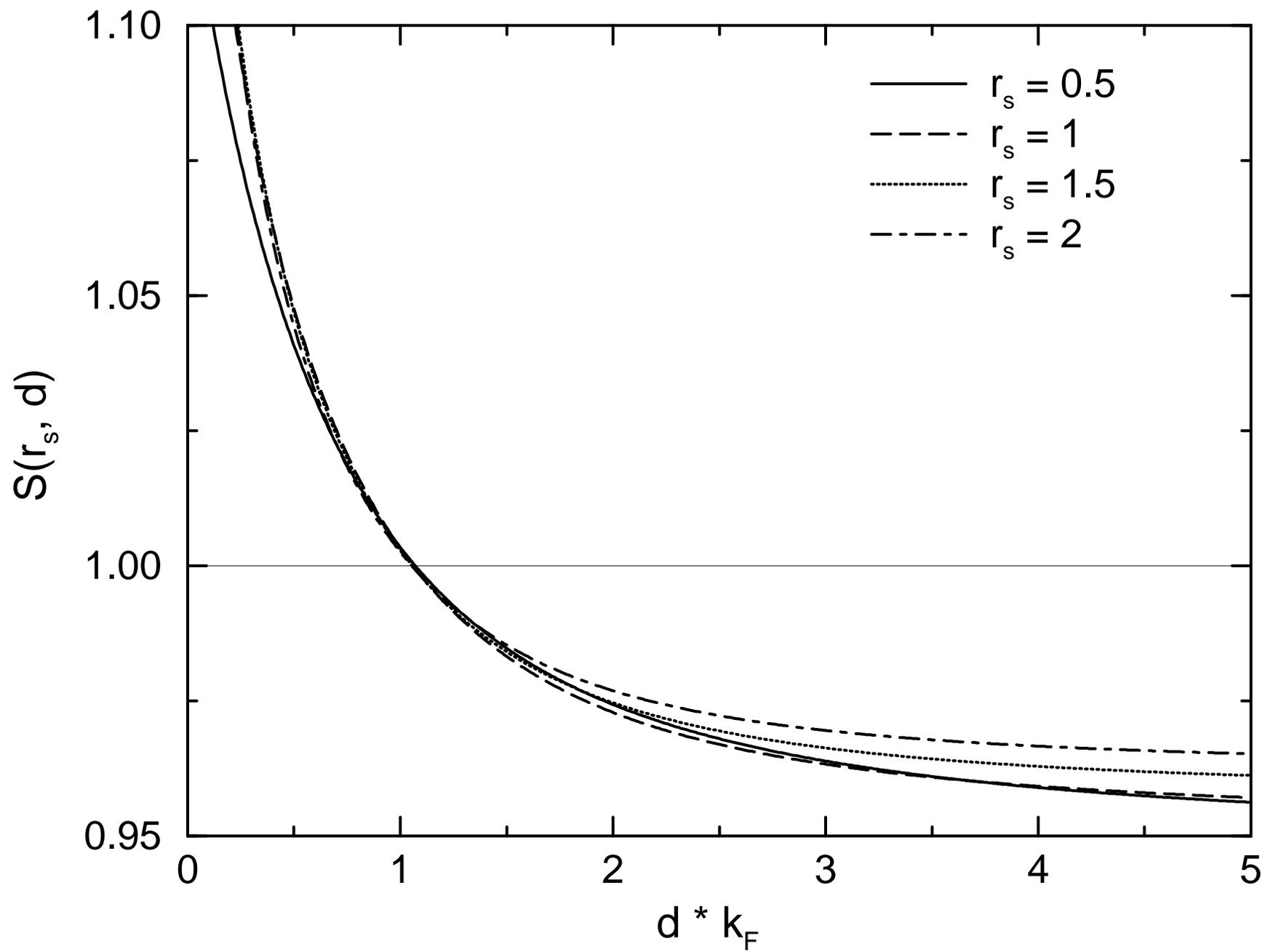
\begin{figure}
\caption{GRPA transverse pseudospin susceptibility Stoner enhancement parameter 
$S$, calculated using statically screened Coulomb interactions, as a function
of $k_{F} d$ for a series of 2D electron gas density parameters. Results
are plotted for $r_{s} = 0.5, 1.0, 1.5, \hbox{ and } 2.0$.}
\label{fig:two}
\end{figure}

The GRPA estimate of $S$ at general $d$ can be
improved by using a static screening approximation for $V_{A}(\vec{q})$
and $V_{E}(\vec{q})$; this procedure is somewhat {\it ad hoc} but it is
known to work reasonably well numerically for analogous
quantities  in single layer systems.  In particular, quantum Monte Carlo 
estimates\cite{mccalc} for the quasiparticle mass of 2D electron systems
are close to screened Hartree-Fock values.  However the reduction of 
the ordinary spin-susceptibility compared to the GRPA is overestimated 
by this approximation.  For $q < 2 k_{F}$, the RPA screened
intralayer and interlayer interactions are\cite{refscreen}
\begin{equation}
V_{A}^{sc}(q) = 2\pi e^{2} \frac{q + k_{TF}[1 - \exp{(-2qd)}]}{q^{2} + 
2q k_{TF} + k_{TF}^{2} [1 - \exp{(-2qd)}]}
\label{eq:16}
\end{equation}
and
\begin{equation}
V_{E}^{sc}(q)  = 2\pi e^{2} \frac{q\exp{(-qd)}}{q^{2} + 2q k_{TF} +
k_{TF}^{2} [1 -\exp{(-2qd)]}}
\label{eq:17}
\end{equation}
Results obtained for the $d$ dependence of the Stoner factor using these
screened interactions are illustrated in Fig.~\ref{fig:two}.  
The results are only weakly dependent on density over this wide
range which covers the density range of typical samples.  (For $r_s \to 0$,
$S \to 1$, independent of $k_F d$.)  Again we find that the 
susceptibility is suppressed rather than enhanced for $k_F d > 1.1$.

As mentioned above, many-body effects in double-layer quantum Hall systems
are often quite sensitive to the precise numerical value of the interlayer
hopping parameter $t$, especially in the limit of strong magnetic
fields.   Reliable comparison between theory and  experiment can be
dependent on accurate estimates of this system parameter.   In this work
we have pointed out that a non-interacting electron interpretation of weak
field  magnetic oscillation experiments does not necessarily yield an
accurate value for $t$. The true value of $t$ can be obtained from the
apparent value by dividing by the Stoner enhancement factor for the transverse
pseudospin susceptibility.  Over the density range of typical samples 
our results are accurately interpolated by the formula
\begin{equation}
S \approx \frac{ 1.19 + 2.37 x + 0.84 x^2}{ 1 + 2.5 x + 0.88 x^2 }
\label{eq:18}
\end{equation}
where $x = k_F d$.  The precise value of $S$ will depend on density,
and on finite thickness effects in the 2D electron layers, and may 
differ somewhat from the GRPA static-screening value.

This work was supported in part by the National Science Foundation under
grant DMR-9416906 and by grant from the Australian Research Council.  
The authors acknowledge helpful interactions with  Greg Boebinger, 
Jim Eisenstein, Zbyszek Gortel, Charles Hanna, and Mansour Shayegan.

\end{document}